\documentclass[12pt]{article}
\usepackage{amssymb}
\usepackage{latexsym}
\usepackage{amsmath}
\usepackage{graphicx}

\def\th{\theta}

\title{\LARGE{Tunnelling of scalar and Dirac particles from squashed charged rotating Kaluza-Klein black holes }}
\author{M. M. Stetsko\footnote{E-mail: mstetsko@gmail.com, mykola@ktf.franko.lviv.ua}\
\\
  {\small Department of Theoretical Physics, Ivan Franko National University of Lviv,}\\
{\small 12 Drahomanov Str., Lviv, UA-79005, Ukraine
         }}

\begin{document}
\maketitle

\abstract{Thermal radiation of scalar particles and Dirac fermions
from squashed charged rotating five-dimensional black holes is
considered. To obtain temperature of the black holes we use the
tunnelling method. In case of scalar particles we make use of the
Hamilton-Jacobi equation. To consider tunnelling of fermions the
Dirac equation was investigated. The examination shows that radial
parts of the action for scalar particles and fermions in
quasi-classical limit in the vicinity of horizon are almost the
same and as a consequence it gives rise to the identical
expressions for the temperature in both cases.}

\section{Introduction}
Hawking radiation has been investigated since the early 70-ies of
the last century \cite{Hawking_CMP75,Brout_PRept95}. To find
temperature of black holes different methods were used.
Semiclassical tunnelling method proposed by Kraus and Wilczek
\cite{Kraus_Wilczek,Kraus_Wilczek_2} and developed in works of
Parikh and Wilczek \cite{Parikh_PRL00} has been attracted a lot of
interest recently. We remark that nowadays the tunnelling method
is comprised of two different approaches namely the approach
proposed by Parikh and Wilczek and it is called the null-geodesic
method and the second one that is known as the Hamilton-Jacobi
method \cite{Padmanabhan_PRD99,angheben_JHEP05}. It is worth
noting that null-geodesic method is truly semiclassical because it
is based on the equation of motion of a classical particle that
moves along a null-geodesic. The Hamilton-Jacobi method is based
on equations that we deal with in quantum theory such as the
Klein-Gordon equation for scalar particles and the Dirac equation
for fermions with spin $s=1/2$. For particles of higher spins
Rarita-Schwinger or Proca equations can be used. So here one
departs from quantum equation of motion and then considers
quasi-classical limit. Despite the different starting points two
methods give rise to the same temperatures if the black hole's
metrics is the same.

These approaches were applied to numerous examples of black holes
and obtained results were in agreement with expressions for
temperature calculated by other methods. Among them we distinguish
Kerr and Kerr-Newman black holes' space-time
\cite{Q_Jiang_PRD06,J_Zhang_PLB06,Chen_PLB08,Kerner_PLB08,Stetsko_EJPC2014},
Taub-NUT space-time \cite{kerner_PRD07} G\"{o}del space-time
\cite{kerner_PRD07_2}, BTZ black holes
\cite{S_Wu_JHEP06,Li_PLB08}, dynamical black holes
\cite{di_Criscienzo_PLB07}, black holes in Ho\v{r}ava-Lifshitz
gravity \cite{Chen_PLB09, Liu_CQG11}, accelerating and rotating
black hole \cite{Rehnam_JCAP11, Sharif_EPJC12}, rotating black
strings \cite{Ahmed_JCAP11} and many others. The review of
tunnelling method was considered in paper \cite{Vanzo_CQG11} where
further references on that subject are given.

Multidimensional black holes have been attracted great attention
in recent years \cite{Emparan_LivRevRel08}. There are great
expectations that multidimensional micro black holes can appear in
high energy particles collisions which are based on the fact that
extra spacelike dimensions can lower the Planck scale up to the
$TeV$-energy region \cite{arkani-hamed,Randall99}. Squashed
Kaluza-Klein black holes represent one of the most interesting
solutions among higher dimensional black holes. The first
five-dimensional Kaluza-Klein black hole's solution was found by
Dobiasch and Maison \cite{Dobiasch_GRG82} and that metrics was
studied in \cite{Gibbons_AnnPhys87}. The solution was generalized
to the five-dimensional Einstein-Maxwell theory
\cite{Ishihara_PTP2006}. To find the solution so called squashing
transformation was used \cite{Ishihara_PTP2006}. The squashing
transformation was applied to construct rotating black hole
solution \cite{Wang_NPB2006}, charged rotating black hole
\cite{NakagawaPRD2008}, charged rotating black hole in G\"{o}del
universe \cite{Steltea_PRD_2008,Matsuno_PRD2008}. A review of
Kaluza-Klein black holes' solutions can be found in
\cite{Tomizawa_PTP2011}.

Different aspects of Kaluza-Klein black holes were also
considered, namely thermodynamics
\cite{Cai_PLB2006,KuritaCQG2007,KuritaCQG2008}, Hawking radiation
and tunnelling method
\cite{Ishihara_prd2007,S_Chen_PRD_2008,S_W_wei_EPJC,D_Y_Chen_CQG_2008,H_l_Li_EPL2010,L_H_Ling_chinPhB2011,Matsuno_PRD11},
quasinormal modes and stabilities
\cite{Ishihara_etal_PRD2008,X_He_PLB2008,x_He_PRD2009,M_Kimura_PRD2008,R_Nishikawa_CQG2010},
geodetic precession \cite{K_Matsuno_PRD2009}, gravitational
lensing \cite{Y_Liu_PRD2010}.

In our work we consider scalar particles and fermion tunnelling
for charged rotating Kaluza-Klein black holes. The paper is
organized as follows. In the second section we investigate
tunnelling of scalar particles. In the third section the
tunnelling of Dirac fermions is examined. The forth section
contains some conclusions.

\section{Tunnelling of scalar particles}
Let us consider tunnelling of scalar particles. The Klein-Gordon
equation for charged massive particles can be written in the form:
\begin{equation}\label{KG_eq}
\frac{1}{\sqrt{-g}}\left(\partial_{\mu}-ieA_{\mu}\right)\sqrt{-g}g^{\mu\nu}
\left(\partial_{\nu}-ieA_{\nu}\right)\Psi-\tilde{m}^2\Psi=0.
\end{equation}
The tunnelling process is supposed to be quasi-classical. To
consider it we assume that quasi-classical wave function takes
form:
\begin{equation}
\Psi=C\exp \left\{\frac{i}{\hbar}I_{\uparrow}\right\}
\end{equation}
Where $I_{\uparrow}=I_{\uparrow}(t,r,\theta,\phi,\psi)$ denotes a
quasi-classical action of emitted particles. In the first order
approximation the Klein-Gordon equation (\ref{KG_eq}) leads to a
Hamilton-Jacobi equation for the relativistic particle:
\begin{equation}\label{HJ_eq_1}
g^{\mu\nu}\left(\partial_{\mu}I_{\uparrow}\partial_{\nu}I_{\uparrow}+e^2A_{\mu}A_{\nu}-
2eA_{\mu}\partial_{\nu}I_{\uparrow}\right)+\tilde{m}^2=0.
\end{equation}
We will examine tunnelling of particles through the horizon of
squashed charged rotating Kaluza-Klein black hole whose metric and
gauge potential were obtained in \cite{NakagawaPRD2008}:
\begin{eqnarray}\label{Sq_KK_BH}
ds^2=-\frac{w(r)}{h(r)}dt^2+\frac{k^2(r)}{w(r)}dr^2+\frac{r^2}{4}
\left[k(r)(\sigma^2_1+\sigma^2_2)+h(r) (f(r)dt+\sigma_3)^2\right]
\end{eqnarray}
where functions $w(r)$, $h(r)$, $k(r)$ and $f(r)$ are defined as
follows
\begin{equation}
w(r)=\frac{(r^2+q)^2-2(m+q)(r^2-a^2)}{r^4}=\frac{(r^2-r^2_+)(r^2-r^2_-)}{r^4}
\end{equation}
\begin{equation}
h(r)=1-\frac{a^2q^2}{r^6}+\frac{2a^2(m+q)}{r^4}
\end{equation}
\begin{equation}
k(r)=\frac{(r^2_{\infty}+q)^2-2(m+q)(r^2_{\infty}-a^2)}{(r^2_{\infty}-r^2)^2}=
\frac{(r^2_{\infty}-r^2_+)(r^2_{\infty}-r^2_-)}{(r^2_{\infty}-r^2)^2}
\end{equation}
\begin{equation}
f(r)=-\frac{2a}{r^2h(r)}\left(\frac{2m+q}{r^2}-\frac{q^2}{r^4}\right)
\end{equation}
Gauge potential is defined by the relation:
\begin{equation}
A=\frac{\sqrt{3}q}{2r^2}\left(dt-\frac{a}{2}\sigma_3\right)
\end{equation}
and left-invariant $1$-forms on $S^3$ are given by
\begin{equation}
\sigma_1=\cos{\psi}d\theta+\sin{\psi}\sin{\theta}d\phi
\end{equation}
\begin{equation}
\sigma_2=-\sin{\psi}d\theta+\cos{\psi}\sin{\theta}d\phi
\end{equation}
\begin{equation}
\sigma_1=d\psi+\cos{\theta}d\phi
\end{equation}
The coordinates $(t,r,\theta,\phi,\psi)$ run the ranges of
$-\infty<t<+\infty$, $0<r<r_{\infty}$,
$0\leqslant\theta\leqslant\pi$, $0\leqslant\phi<2\pi$,
$0\leqslant\psi<4\pi$ respectively. The case $k(r)=1$ or
equivalently $r_{\infty}\rightarrow\infty$ leads to the Cveti\v{c}
solution \cite{CveticNpb,CveticPLB04}. The parameters $a$, $m$,
$q$ and $r_{\infty}$ fulfil the inequalities:
\begin{eqnarray}
m>0\label{m_1},\\ q^2+2(m+q)a^2>0 \label{maq_1},\\
(r^2_{\infty}+q)^2-2(m+q)(r^2_{\infty}-a^2)>0,\label{mqa_rinf_1}\\
(m+q)(m-q-2a^2)>0\label{mqa_2}, \\ m+q>0\label{mq_2}.
\end{eqnarray}
It was pointed out by the authors \cite{NakagawaPRD2008} that
inequalities (\ref{m_1})-(\ref{mqa_2}) are necessary for the
existence of two horizons and the last inequality (\ref{mq_2})
provides the absence of closed timelike curves outside the outer
horizon. The horizons are defined by the equation $w(r)=0$ and it
leads to: $r^2_+=m+\sqrt{(m+q)(m-q-2a^2)}$ and
$r^2_-=m-\sqrt{(m+q)(m-q-2a^2)}$. The metric (\ref{Sq_KK_BH}) also
diverges at $r=r_{\infty}$ but it is an apparent singularity
corresponding to the spatial infinity. To prove it one should make
a coordinate transformation
\begin{equation}
\rho=\rho_0\frac{r^2}{r^2_{\infty}-r^2}
\end{equation}
where $\rho_0$ is defined by
\begin{equation}
\rho^2_0=\frac{(r^2_{\infty}+q)^2-2(m+q)(r^2-a^2)}{4r^2_{\infty}}=
\frac{(r^2_{\infty}-r^2_+)(r^2_{\infty}-r^2_-)}{4r^2_{\infty}}
\end{equation}
It is clear when $r\rightarrow r_{\infty}$ then $\rho\rightarrow
\infty$. It was also shown \cite{NakagawaPRD2008} that asymptotic
time (time for a distant observer) differs from the coordinate one
and takes form:
\begin{equation}\label{asymp_time_sqKK}
\tilde{t}=\frac{2r^2_{\infty}\rho_0}{\sqrt{r^6_{\infty}-a^2(q^2-2(m+q)r^2_{\infty})}}t=\beta
t
\end{equation}

Now we consider the Hamilton-Jacobi equation (\ref{HJ_eq_1}) in
case of the black hole's metric (\ref{Sq_KK_BH}). The functions
$w(r), k(r)$, $h(r)$ and $f(r)$ depend only on the radial variable
$r$ and it gives rise to the conclusion that the angular variables
can be separated form the radial one (or at least some of them).
Using this fact we suppose that the angular variables $\phi$ and
$\psi$ can be separated. So the action for emitted particle
$I_{\uparrow}$ might be written in the form
\begin{equation}\label{action_sc}
I_{\uparrow}=-E\tilde{t}+W(r,\theta)+J\phi+L\psi
\end{equation}
where $J$ and $L$ are constants and $\tilde{t}$ denotes time for a
distant observer which is defined by the relation
(\ref{asymp_time_sqKK}).

As a result the Hamilton-Jacobi equation can be represented in the
form:
\begin{eqnarray}
\frac{w(r)}{k^2(r)}W^2_r+\frac{4W^2_{\theta}}{r^2k(r)}-\frac{h(r)}{w(r)}\left(\beta
E+f(r)L+\frac{\sqrt{3}qe}{2r^2}\left(1+\frac{af(r)}{2}\right)\right)^2\\\nonumber
+\frac{1}{r^2h(r)}\left(2L+\frac{\sqrt{3}qea}{2r^2}\right)^2+\frac{4}{r^2k(r)}\left(\frac{J}{\sin{\theta}}-L\cot{\theta}\right)^2
+\tilde{m}^2=0
\end{eqnarray}
One can see that variables $r$ and $\theta$ can be separated in
the same way. So the function $W(r,\theta)$ can be written as
follows:
\begin{equation}\label{sep_r_theta}
W(r,\theta)=R(r)+\Theta(\theta)
\end{equation}
Then for radial part we obtain:
\begin{eqnarray}\label{rad_der_sc_SqKK}
R'(r)=\frac{\beta k(r)\sqrt{h(r)}}{w(r)}\left[\left(
E+\frac{f(r)}{\beta}L+\frac{\sqrt{3}qe}{2\beta
r^2}\left(1+\frac{af(r)}{2}\right)\right)^2\right.\\\nonumber
\left.-\frac{w(r)}{\beta^2h(r)}\left(\frac{1}{r^2h(r)}\left(2L+\frac{\sqrt{3}qea}{2r^2}\right)^2
+\frac{4D}{r^2k(r)}+\tilde{m}^2\right)\right]^{1/2}
\end{eqnarray}
where $D$ is a constant that appeared after separation of
variables.

In the vicinity of the horizon point $w(r_+)=0$ one should use the
decomposition $w(r)=w'(r_+)(r-r_+)=2(r^2_+-r^2_-)(r-r_+)/r^3_+$.
Then the latter equation can be written in the form:
\begin{equation}
R'(r)=\frac{\beta}{2}\frac{r^2_{\infty}-r^2_-}{r^2_{\infty}-r^2_+}\frac{\sqrt{r^6_+-a^2q^2-2a^2(m+q)r^2_+}}
{(r^2_+-r^2_-)(r-r_+)}\left(
E+\frac{f(r_+)L}{\beta}+\frac{\sqrt{3}qe}{2\beta
r^2_+}\left(1+\frac{af(r_+)}{2}\right)\right)
\end{equation}
Inegrating this expression in the vicinity of the horizon point we
obtain:
\begin{equation}\label{radial_part_sc}
R(r)=\frac{\beta}{2}\frac{r^2_{\infty}-r^2_-}{r^2_{\infty}-r^2_+}\frac{\sqrt{r^6_+-a^2q^2-2a^2(m+q)r^2_+}}
{(r^2_+-r^2_-)}\left(
E+\frac{f(r_+)L}{\beta}+\frac{\sqrt{3}qe}{2\beta
r^2_+}\left(1+\frac{af(r_+)}{2}\right)\right)\int^{r_++\varepsilon}_{r_+-\varepsilon}\frac{dr}{(r-r_+)}
\end{equation}
We consider the tunnelling process and it means that the action
for the emitted particles $I_{\uparrow}$ (\ref{action_sc}) takes
complex values. This fact immediately follows from the form of the
integral for the radial part of the action (\ref{radial_part_sc}).
The function we integrate in (\ref{radial_part_sc}) has a pole at
the point $r_+$. Then we should integrate around the pole and as a
result the complex values appear. To obtain temperature of the
black hole we take into account only the radial part of the action
(\ref{radial_part_sc}). So we can write:
\begin{equation}\label{Im_out_sc}
Im
R_{\uparrow}=\pi\frac{\beta}{2}\frac{r^2_{\infty}-r^2_-}{r^2_{\infty}-r^2_+}\frac{\sqrt{r^6_+-a^2q^2-2a^2(m+q)r^2_+}}
{(r^2_+-r^2_-)}\left(
E+\frac{f(r_+)L}{\beta}+\frac{\sqrt{3}qe}{2\beta
r^2_+}\left(1+\frac{af(r_+)}{2}\right)\right)
\end{equation}
It was supposed \cite{Padmanabhan_PRD99,Vanzo_CQG11} that
probabilities of crossing a black hole's horizon can be defined
as:
\begin{equation}\label{cross_prob}
P_{out}\propto\exp\{-2ImR_{\uparrow}\}, \quad
P_{int}\propto\exp\{-2ImR_{\downarrow}\}
\end{equation}
The ratio for these two probabilities leads to the Boltzmann
factor $\exp\{-E/T\}$ which shows that radiation is thermal
\cite{Hartle_PRD76}:
\begin{equation}\label{tunn_prob}
\Gamma=\frac{P_{out}}{P_{in}}=\frac{\exp\{-2ImR_{\uparrow}\}}{\exp\{-2ImR_{\downarrow}\}}=\exp\left\{-\frac{E}{T}\right\}.
\end{equation}
The imaginary part of the radial part of the action for ingoing
particles can be calculated in the same way as it was carried out
for the outgoing case. The resulting imaginary part will take only
the opposite sign in comparison to the expression
(\ref{Im_out_sc}), so $ImR_{\downarrow}=-ImR_{\uparrow}$.
Substituting the expression (\ref{radial_part_sc}) and taking into
consideration mentioned above remark we arrive at:
\begin{equation}
\Gamma=\exp\left\{-2\pi\beta\frac{r^2_{\infty}-r^2_-}{r^2_{\infty}-r^2_+}\frac{\sqrt{r^6_+-a^2q^2-2a^2(m+q)r^2_+}}
{(r^2_+-r^2_-)}\left(
E+\frac{f(r_+)L}{\beta}+\frac{\sqrt{3}qe}{2\beta
r^2_+}\left(1+\frac{af(r_+)}{2}\right)\right)\right\}
\end{equation}\label{Temp_Sq_kk}
Having supposed that emission of particles is thermal we obtain
the temperature of the black hole:
\begin{equation}\label{temper_Sq_KK}
T=\frac{1}{2\pi}\frac{r^2_+-r^2_-}{r_{\infty}(r^2_{\infty}-r^2_-)}\sqrt{\frac{r^2_{\infty}-r^2_+}{r^2_{\infty}-r^2_-}}
\sqrt{\frac{r^6_{\infty}-a^2(q^2-2(m+q)r^2_{\infty})}{r^6_+-a^2(q^2-2(m+q)r^2_+)}}
\end{equation}
The result we obtain is in agreement with the expression for the
surface gravity that was found in \cite{NakagawaPRD2008}.

Now we consider a squashed black hole in G\"{o}del universe. The
metric of the black hole takes the following form
\cite{Steltea_PRD_2008}:
\begin{equation}\label{goedel_sq_bh}
ds^2=-k(r)dt^2-2g(r)\sigma_3dt+h(r)\sigma_3^2+\frac{\chi^2(r)}{V(r)}dr^2+\frac{r^2}{4}
\left[\chi(r)(\sigma^2_1+\sigma^2_2)\right]
\end{equation}
where
\begin{eqnarray}
k(r)=1-\frac{2m}{r^2}+\frac{q^2}{r^4}
\\ g(r)=jr^2+3jq+\frac{(2m-q)a}{2r^2}-\frac{q^2a}{2r^4}
\\
h(r)=\frac{r^2}{4}-j^2r^2(r^2+2m+6q)+3jqa+\frac{(m-q)a^2}{2r^2}-\frac{q^2a^2}{4r^4};
\\
V(r)=1+\frac{8j(m+q)(a+2j(m+2q))-2m}{r^2}+\frac{2(m-q)a^2+q^2[1-16ja-8j^2(m+3q)]}{r^4};
\\\chi(r)=\frac{c^2+2c(m-4j(m+q)[a+2j(m+2q)])+q^2+2a^2(m-q)-8q^2j[2a+j(m+3q)]}{(r^2+c)^2}
\end{eqnarray}
We note that here we keep notations given in the paper
\cite{Steltea_PRD_2008}. The squashing functions here is
$\chi(r)$. It was shown that in case when G\"{o}del parameter
$j=0$ we arrive at the previously considered metric
(\ref{Sq_KK_BH}). We also note that the squashing parameter $c$
has to be chosen negative $c=-r^2_0$ (the parameter $r_0$
corresponds to the parameter $r_{\infty}$ of previously considered
metric) The electromagnetic one-form potential can be written as
follows:
\begin{equation}
A=\left(\frac{\sqrt{3}q}{2r^2}-\Phi\right)dt+\frac{\sqrt{3}}{2}\left(jr^2+2jq-\frac{aq}{2r^2}\right)\sigma_3
\end{equation}
and here $\Phi$ is a constant which one can find from the
requirement that electromagnetic potential should be regular at
the horizon. Horizon points for the black hole given by the metric
(\ref{goedel_sq_bh}) can be found as roots of equation $V(r)=0$
\cite{Steltea_PRD_2008}:
\begin{eqnarray}
r^2_{\pm}=m-4j(m+q)(a+2j(m+2q))\pm\sqrt{\delta},
\\\delta=(m-q-8j^2(m+q)^2)[m+q-2a^2-8ja(m+2q)-8j^2(m+2q)^2]
\end{eqnarray}
We note that in the limit $j=0$ we recover the previously
considered black hole's solution (\ref{Sq_KK_BH}).

It should be noticed that black hole solutions in G\"{o}del
universe possess close timelike curves (CTC).  In our case we will
consider situation when $r_-<r_+<r_0<r_{CTC}$ so this peculiarity
of the solution (\ref{goedel_sq_bh}) is not important for us.

Now we examine the Hamilton-Jacobi equation for the squashed
G\"{o}del black hole. Similarly as in the previous case components
 of metric tensor do not depend explicitly on the coordinates $t, \phi,
 \psi$ and it allows us to separate those angular and time
 coordinates. So the action for an emitted particle can be chosen
 in the form (\ref{action_sc}). Having accomplished the separation of
 variables we  write the Hamilton-Jacobi equation:
\begin{eqnarray}
\frac{V(r)}{\chi^2(r)}W^2_r+\frac{4}{r^2\chi(r)}\left(W^2_{\theta}+\frac{(J-L\cos{\theta})^2}{\sin^2{\theta}}\right)+
\frac{4}{r^2V(r)}\left(-h(r\left[E+e\left(\frac{\sqrt{3}q}{2r^2}-\Phi\right)\right]^2+\right.\nonumber\\
\left.2g(r)\left[\left(E+e\left(\frac{\sqrt{3}q}{2r^2}-\Phi\right)\right)\left(L-\frac{\sqrt{3}e}{2}
\left(jr^2+2jq-\frac{aq}{2r^2}\right)\right)\right]\right.\nonumber\\
\left.+k(r)\left(L-\frac{\sqrt{3}e}{2}\left(jr^2+2jq-\frac{aq}{2r^2}\right)\right)^2\right)+\tilde{m}^2=0
\end{eqnarray}
We note that here we have used the relation
$g^2(r)+k(r)h(r)=r^2V(r)/4$ which can be verified easily.
Similarly to the previous case the radial $r$ and angular $\theta$
variables can be separated. So we make use of the relation
(\ref{sep_r_theta}) and write the relation for the derivative of
radial part of action $I_{\uparrow}$
\begin{eqnarray}
R'(r)=\frac{2\chi(r)\sqrt{h(r)}}{rV(r)}\left(\left[E+e\left(\frac{\sqrt{3}q}{2r^2}-
\Phi\right)-\frac{g(r)}{h(r)}\left(L-\frac{\sqrt{3}e}{2}\left(jr^2+2jq-\frac{aq}{2r^2}\right)\right)\right]^2-
\right.\nonumber\\
\left.
\frac{r^2V(r)}{h(r)}\left[\frac{1}{h(r)}\left(L-\frac{\sqrt{3}e}{2}\left(jr^2+2jq-\frac{aq}{2r^2}\right)\right)^2
+\frac{4D}{r^2\chi(r)}+\tilde{m}^2\right]\right)^{1/2}
\end{eqnarray}
Integrating the obtained relation around the horizon point $r_+$
and taking the imaginary part of it we arrive at:
\begin{eqnarray}\label{imag_part_goedel}
ImR_{\uparrow}=\pi\frac{r^2_+\chi(r_+)\sqrt{h(r_+)}}{(r^2_+-r^2_-)}\left[E+e\left(\frac{\sqrt{3}q}{2r^2_+}-
\Phi\right)-\frac{g(r_+)}{h(r_+)}\left(L-\frac{\sqrt{3}e}{2}\left(jr^2_++2jq-\frac{aq}{2r^2_+}\right)\right)\right]
\end{eqnarray}
To find temperature of the black hole we make use of the same
procedure as we have done earlier. It should be emphasized that
similarly to the previous metric to obtain correct relation for
temperature we have to take into account the fact that a distant
observer uses asymptotic time but not a coordinate one. So we
perform transformation similar to (\ref{asymp_time_sqKK})
$t\rightarrow t/N_0$, where parameter $N_0$ takes form:
\begin{equation}
N^2_0=\frac{r^2_0V(r_0)}{4h(r_{0})}.
\end{equation}
This transformation leads to the replacement $E\rightarrow EN_0$
in the right hand side of the relation (\ref{imag_part_goedel}).
We note that in case of previously considered metric the situation
was the same.

So we write:
\begin{equation}\label{temp_Godel_sq}
T=\frac{1}{2\pi}\frac{r_{0}}{r^2_+}\frac{r^2_+-r^2_-}{r^2_{0}-r^2_-}
\sqrt{\frac{r^2_{0}-r^2_+}{r^2_{0}-r^2_-}}\sqrt{\frac{h(r_{0})}{h(r_+)}}
\end{equation}
In the limit when the G\"{o}del parameter is equal to zero ($j=0$)
the obtained relation (\ref{temp_Godel_sq}) can be represented in
the form (\ref{temper_Sq_KK}).

\section{Tunnelling of a charged spin-$1/2$ particle from squashed Kaluza-Klein black hole}
In this section we examine the tunnelling method for Dirac
particles. For the first time tunnelling method for fermions was
considered by Kerner and Mann \cite{Kerner_PLB08}. Then it was
successfully applied to the vast area of black holes
\cite{Vanzo_CQG11}. In case of scalar particles the starting point
was the Klein-Gordon equation (\ref{KG_eq}) which leads to the
Hamilton-Jacobi equation (\ref{HJ_eq_1}) in the quasi-classical
limit. To investigate the tunnelling of fermions the Klein-Gordon
equation should be replaced by the Dirac equation. Then similarly
to the case of scalar particles quasi-classical limit should be
taken. Firstly we consider black hole's metric (\ref{Sq_KK_BH})
and then go to a bit more general metric of the squashed black
hole in G\"{o}del universe (\ref{goedel_sq_bh}). The Dirac
equation for an electrically charged particle takes form:
\begin{equation}\label{Dirac_General}
i\gamma^{\mu}\left(D_{\mu}-\frac{ie}{\hbar}A_{\mu}\right)\Psi+\frac{\tilde{m}}{\hbar}\Psi=0
\end{equation}
where $D_{\mu}=\partial_{\mu}+\Omega_{\mu}$,
$\Omega_{\mu}=\frac{1}{8}\Gamma^{\alpha\beta}_{\mu}[\gamma^{\beta},\gamma^{\alpha}]$
and $\gamma^{\mu}$ matrices fulfil commutation relation:
\begin{equation}
\{\gamma^{\mu},\gamma^{\nu}\}=2g^{\mu\nu}\hat{1}
\end{equation}
Matrices $\gamma^{\mu}$ can be defined in different manners and in
our work we take them in the following form:
\begin{eqnarray}
\hat{\gamma}^{\tilde{t}}=\beta\sqrt{\frac{h(r)}{w(r)}}\hat{\gamma}^{0},
\quad
\hat{\gamma}^{r}=\frac{\sqrt{w(r)}}{k(r)}\hat{\gamma}^{3},\quad
\hat{\gamma}^{\theta}=\frac{2}{r\sqrt{k(r)}}\hat{\gamma}^{1},
\\\nonumber
\hat{\gamma}^{\phi}=\frac{2}{r\sqrt{k(r)}\sin{\theta}}\hat{\gamma}^{2},\quad
\hat{\gamma}^{\psi}=-f(r)\sqrt{\frac{h(r)}{w(r)}}\hat{\gamma}^{0}-\frac{2\cot{\theta}}{r\sqrt{k(r)}}\hat{\gamma}^{2}+
\frac{2}{r\sqrt{h(r)}}\hat{\gamma}^{4}
\end{eqnarray}
It should be noted that we consider black hole's metric
(\ref{Sq_KK_BH}). Matrices $\gamma^A$ take form:
\begin{eqnarray}
\gamma^0=
\begin{pmatrix}
0 & I \\
-I & 0\\
\end{pmatrix} \quad
\gamma^1=
\begin{pmatrix}
0 & \sigma_1 \\
\sigma_1 & 0\\
\end{pmatrix}\\
\gamma^2=
\begin{pmatrix}
0 & \sigma_2 \\
\sigma_2 & 0\\
\end{pmatrix} \quad
\gamma^3=
\begin{pmatrix}
0 & \sigma_3 \\
\sigma_3 & 0\\
\end{pmatrix}\quad
\gamma^4=
\begin{pmatrix}
-I & 0 \\
0 & -I\\
\end{pmatrix}
\end{eqnarray}
and $\sigma_i$ are the Pauli matrices.

So the Dirac equation can be written in the form:
\begin{eqnarray}\label{dirac_sq_kk_1}
i\left[\beta\sqrt{\frac{h(r)}{w(r)}}\hat{\gamma}^0\partial_{\tilde{t}}+\frac{\sqrt{w(r)}}{k(r)}
\hat{\gamma}^3\partial_{r}+\frac{2}{r\sqrt{k(r)}}\hat{\gamma}^1\partial_{\theta}+\nonumber
\frac{2}{r\sqrt{k(r)}\sin{\theta}}\hat{\gamma}^2\partial_{\phi}\right.\\
\left.+\left(-f(r)\sqrt{\frac{h(r)}{w(r)}}\hat{\gamma}^{0}-\frac{2\cot{\theta}}{r\sqrt{k(r)}}\hat{\gamma}^{2}+
\frac{2}{r\sqrt{h(r)}}\hat{\gamma}^{4}\right)\partial_{\psi}
\right]\Psi+\\\nonumber \frac{\sqrt{3}qe}{2\hbar
r^2}\left[\sqrt{\frac{h(r)}{w(r)}}\left(1+\frac{af(r)}{2}\right)\hat{\gamma}^0-
\frac{a}{r\sqrt{h(r)}}\hat{\gamma}^4\right]\Psi+\frac{\tilde{m}}{\hbar}\Psi=0
\end{eqnarray}
We also remark that here spin connection terms $\Omega_{\mu}$ are
omitted because we will consider quasi-classical limit and take
the lowest order approximation whereas spin connection gives rise
to the terms of the next order.

The wave functions with spin up and down can be chosen in the
form:
\begin{eqnarray}\label{wave_func_up}
\Psi_{\uparrow}=
\begin{pmatrix}
A(t,r,\th,\phi,\psi)\\ 0 \\ B(t,r,\th,\phi,\psi)\\0\\
\end{pmatrix}
\exp{\left(\frac{i}{\hbar}I_{\uparrow}(t,r,\th,\phi,\psi)\right)};\quad
\Psi_{\downarrow}=
\begin{pmatrix}
0 \\C(t,r,\th,\phi,\psi) \\0 \\B(t,r,\th,\phi,\psi)\\
\end{pmatrix}
\exp{\left(\frac{i}{\hbar}I_{\downarrow}(t,r,\th,\phi,\psi)\right)},
\end{eqnarray}
where $I_{\uparrow}$ and $I_{\downarrow}$ are the action for the
Dirac particles with spin-up ($\uparrow$) and spin-down
($\downarrow$) tunnelling through the horizon.

Substituting (\ref{wave_func_up}) into the Dirac equation
(\ref{dirac_sq_kk_1}) and taking the lowest order terms (the terms
proportional to $\hbar^{-1}$) we arrive at
\begin{eqnarray}\label{D_eq_1}
B\left(\sqrt{\frac{h(r)}{w(r)}}\left[-\beta\partial_{\tilde{t}}I_{\uparrow}+f(r)\nonumber
\partial_{\psi}I_{\uparrow}+\frac{\sqrt{3}qe}{2r^2}\left(1+\frac{af(r)}{2}\right)\right]-\right.\\
\left.\frac{\sqrt{w(r)}}{k(r)}\partial_{r}I_{\uparrow}\right)+A\left(\frac{2}{r\sqrt{h(r)}}
\partial_{\psi}I_{\uparrow}+\frac{\sqrt{3}qea}{2r^3\sqrt{h(r)}}+\tilde{m}\right)=0,
\end{eqnarray}
\begin{eqnarray}\label{D_eq_2}
\frac{2B}{r\sqrt{k(r)}}\left(-\partial_{\theta}I_{\uparrow}-\frac{i}{\sin{\theta}}\partial_{\phi}I_{\uparrow}+
i\cot{\theta}\partial_{\psi}I_{\uparrow}\right)=0,
\end{eqnarray}
\begin{eqnarray}\label{D_eq_3}
B\left(\sqrt{\frac{h(r)}{w(r)}}\left[-\beta\partial_{\tilde{t}}I_{\uparrow}+f(r)\nonumber
\partial_{\psi}I_{\uparrow}+\frac{\sqrt{3}qe}{2r^2}\left(1+\frac{af(r)}{2}\right)\right]-\right.\\
\left.\frac{\sqrt{w(r)}}{k(r)}\partial_{r}I_{\uparrow}\right)+A\left(\frac{2}{r\sqrt{h(r)}}
\partial_{\psi}I_{\uparrow}+\frac{\sqrt{3}qea}{2r^3\sqrt{h(r)}}+\tilde{m}\right)=0,
\end{eqnarray}
\begin{eqnarray}\label{D_eq_4}
\frac{2A}{r\sqrt{k(r)}}\left(-\partial_{\theta}I_{\uparrow}-\frac{i}{\sin{\theta}}\partial_{\phi}I_{\uparrow}+
i\cot{\theta}\partial_{\psi}I_{\uparrow}\right)=0.
\end{eqnarray}
It follows immediately from the obtained equations that all the
variables can be separated.  The action $I_{\uparrow}$ is supposed
to take form:
\begin{equation}\label{action_gen_ferm}
I_{\uparrow}=-Et+J\phi+L\psi+R(r)+\Theta(\th),
\end{equation}
where $E$ is the energy of emitted Dirac particles and $J$ and $L$
are angular momenta corresponding to the angles $\phi$ and $\psi$
respectively. We also remark that in case of scalar particles we
initially assumed that just part of variables can be separated and
after little algebra we concluded that we had complete separation
of variables. The complete separation of variables for Dirac
particles follows from the written equations
(\ref{D_eq_1})-(\ref{D_eq_4}). Having inserted the ansatz
(\ref{action_gen_ferm}) into equations
(\ref{D_eq_1})-(\ref{D_eq_4}) we obtain:
\begin{eqnarray}\label{D_eq_1m}
B\left(\sqrt{\frac{h(r)}{w(r)}}\left[\beta
E+f(r)L+\frac{\sqrt{3}qe}{2r^2}\left(1+\frac{af(r)}{2}\right)\right]\right.\nonumber
\\\left.-\frac{\sqrt{w(r)}}{k(r)}R'(r)\right)+
A\left(\frac{1}{r\sqrt{h(r)}}\left(2L+\frac{\sqrt{3}qea}{2r^2}\right)+\tilde{m}\right)=0
\end{eqnarray}
\begin{eqnarray}\label{D_eq_2m}
-\frac{2B}{r\sqrt{k(r)}}\left(\Theta'+\frac{iJ}{\sin{\theta}}-iL\cot{\theta}\right)=0
\end{eqnarray}
\begin{eqnarray}\label{D_eq_3m}
A\left(-\sqrt{\frac{h(r)}{w(r)}}\left[\beta
E+f(r)L+\frac{\sqrt{3}qQe}{2r^2}\left(1+\frac{af(r)}{2}\right)\right]\right.\nonumber
\\\left.-\frac{\sqrt{w(r)}}{k(r)}R'(r)\right)+
A\left(\frac{-1}{r\sqrt{h(r)}}\left(2L+\frac{\sqrt{3}qea}{2r^2}\right)+\tilde{m}\right)=0
\end{eqnarray}
\begin{eqnarray}\label{D_eq_4m}
-\frac{2A}{r\sqrt{k(r)}}\left(\Theta'+\frac{iJ}{\sin{\theta}}-iL\cot{\theta}\right)=0
\end{eqnarray}
System of equations (\ref{D_eq_1m}) and (\ref{D_eq_3m}) has a
nontrivial solution for parameters $A$ and $B$ if and only if the
determinant of corresponding matrix is equal to zero. That
requirement gives rise to the following equation:
\begin{eqnarray}\label{rad_deriv_ferm_SqKK}
R'(r)=\frac{\beta k(r)\sqrt{h(r)}}{w(r)}\left[\left(\nonumber
E+\frac{f(r)}{\beta}L+\frac{\sqrt{3}qe}{2\beta r^2}\left(1+\frac{af(r)}{2}\right)\right)^2\right.\\
\left.-\frac{w(r)}{\beta^2h(r)}\left(\frac{1}{r^2h(r)}\left(2L+\frac{\sqrt{3}qea}{2r^2}\right)^2-\tilde{m}^2\right)\right]^{1/2}
\end{eqnarray}
The structure of the obtained relation is similar to the relation
for scalar particles (\ref{rad_der_sc_SqKK}) and it is not
accidental coincidence because we consider quasi-classical
approximation for both types of particles. In the vicinity of the
horizon point we again use the decomposition
$w(r)=w'(r_+)(r-r_+)=2(r^2_+-r^2_-)(r-r_+)/r^3_+$. So we can
write:
\begin{equation}
R'(r)=\frac{\beta}{2}\frac{r^2_{\infty}-r^2_-}{r^2_{\infty}-r^2_+}\frac{\sqrt{r^6_+-a^2q^2-2a^2(m+q)r^2_+}}
{(r^2_+-r^2_-)(r-r_+)}\left(
E+\frac{f(r_+)L}{\beta}+\frac{\sqrt{3}qe}{2\beta
r^2_+}\left(1+\frac{af(r_+)}{2}\right)\right)
\end{equation}
To obtain temperature for Dirac particles we follow the same steps
that we have made for scalar particles. The similarity of the
obtained relations for the derivatives of radial part of the
action brings us to the conclusion that temperature for tunnelling
fermions has to be the same as for scalar particles and will be
defined by the expression (\ref{temper_Sq_KK}). We also note that
equality of temperatures for scalar particles and Dirac fermions
follows from the fact that both types of particles are examined
quasi-classically.

Now we proceed to the tunnelling of Dirac particles in case of
black hole's metric (\ref{goedel_sq_bh}). We note that fermion
tunnelling  from the squashed black hole in G\"{o}del universe was
investigated in \cite{L_H_Ling_chinPhB2011} but the author
considered just particular case of nonrotating black hole without
charge. To write the Dirac equation gamma matrices should be
defined. The gamma matrices take the form:
\begin{eqnarray}
\hat{\gamma}^{{t}}=\frac{2}{r}\sqrt{\frac{h(r)}{V(r)}}\hat{\gamma}^{0},
\quad
\hat{\gamma}^{r}=\frac{\sqrt{V(r)}}{\chi(r)}\hat{\gamma}^{3},\quad
\hat{\gamma}^{\theta}=\frac{2}{r\sqrt{\chi(r)}}\hat{\gamma}^{1},
\\\nonumber
\hat{\gamma}^{\phi}=\frac{2}{r\sqrt{\chi(r)}\sin{\theta}}\hat{\gamma}^{2},\quad
\hat{\gamma}^{\psi}=\frac{2g(r)}{r\sqrt{h(r)V(r)}}\hat{\gamma}^{0}-\frac{2\cot{\theta}}{r\sqrt{k(r)}}\hat{\gamma}^{2}+
\frac{1}{r\sqrt{h(r)}}\hat{\gamma}^{4}
\end{eqnarray}
The Dirac equation can be written as follows:
\begin{eqnarray}\label{Dirac_goedel}
i\left(\frac{2\sqrt{h(r)}}{r\sqrt{V(r)}}\hat{\gamma}^0\partial_t+\frac{\sqrt{V(r)}}{\chi(r)}
\hat{\gamma}^3\partial_r+\frac{2}{r\sqrt{\chi(r)}}\hat{\gamma}^1\partial_{\theta}+
\frac{2}{r\sqrt{\chi(r)}\sin{\theta}}\hat{\gamma}^2\partial_{\varphi}+\right.\nonumber\\
\left.\left[\frac{2g(r)}{r\sqrt{h(r)V(r)}}\hat{\gamma}^{0}-\frac{2\cot{\theta}}{r\sqrt{k(r)}}\hat{\gamma}^{2}+
\frac{1}{\sqrt{h(r)}}\hat{\gamma}^{4}\right]\partial_{\psi}\right)\Psi+
\frac{e}{\hbar}\left(\frac{2\sqrt{h(r)}}{r\sqrt{V(r)}}\hat{\gamma}^0\times\right.\nonumber\\
\left.\left(\frac{\sqrt{3}q}{2r^2}-\Phi\right)+\frac{\sqrt{3}\cos{\theta}}{r\sqrt{\chi(r)}
\sin{\theta}}\hat{\gamma}^{2}\left(jr^2+2jq-\frac{aq}{2r^2}\right)+\left[\frac{2g(r)}
{r\sqrt{h(r)V(r)}}\hat{\gamma}^{0}-\right.\right.\nonumber\\
\left.\left.\frac{2\cot{\theta}}{r\sqrt{k(r)}}\hat{\gamma}^{2}+
\frac{1}{\sqrt{h(r)}}\hat{\gamma}^{4}\right]\frac{\sqrt{3}}{2}\left(jr^2+2jq-\frac{aq}{2r^2}\right)\right)\Psi
+\frac{m}{\hbar}\Psi=0
\end{eqnarray}
Here we also omitted spin connection terms.

Now we suppose that the wavefunction takes the same form as in the
previous case: (\ref{wave_func_up}). Having substituted the
wavefunction (\ref{wave_func_up}) into the equation
(\ref{Dirac_goedel}) we arrive at:
\begin{eqnarray}\label{Dirac_Goed_1}
B\left[-\frac{2\sqrt{h(r)}}{r\sqrt{V(r)}}\partial_tI_{\uparrow}-\frac{\sqrt{V(r)}}{\chi(r)}
\partial_rI_{\uparrow}-\frac{2g(r)}{r\sqrt{h(r)V(r)}}\partial_{\psi}I_{\uparrow}+\right.\nonumber\\
\left.\frac{2e\sqrt{h(r)}}{r\sqrt{V(r)}}\left(\frac{\sqrt{3}q}{2r^2}-\Phi\right)+
\frac{2eg(r)}{r\sqrt{h(r)V(r)}}\frac{\sqrt{3}}{2}\left(jr^2+2jq-\frac{aq}{2r^2}\right)\right]+\nonumber\\
A\left[\frac{1}{\sqrt{h(r)}}\partial_{\psi}I_{\uparrow}-\frac{1}{\sqrt{h(r)}}\frac{\sqrt{3}}{2}
\left(jr^2+2jq-\frac{aq}{2r^2}\right)+\tilde{m}\right]=0
\end{eqnarray}
\begin{eqnarray}\label{Dirac_Goed_2}
\frac{2B}{r\sqrt{\chi(r)}}\left[-\partial_{\theta}I_{\uparrow}-\frac{i}{\sin{\theta}}
\partial_{\varphi}I_{\uparrow}-i\cot{\theta}\partial_{\psi}I_{\uparrow}\right]=0
\end{eqnarray}
\begin{eqnarray}\label{Dirac_Goed_3}
A\left[\frac{2\sqrt{h(r)}}{r\sqrt{V(r)}}\partial_tI_{\uparrow}-\frac{\sqrt{V(r)}}{\chi(r)}
\partial_rI_{\uparrow}+\frac{2g(r)}{r\sqrt{h(r)V(r)}}\partial_{\psi}I_{\uparrow}-\right.\nonumber\\
\left.\frac{2e\sqrt{h(r)}}{r\sqrt{V(r)}}\left(\frac{\sqrt{3}q}{2r^2}-\Phi\right)-
\frac{\sqrt{3}eg(r)}{r\sqrt{h(r)V(r)}}\left(jr^2+2jq-\frac{aq}{2r^2}\right)\right]+\nonumber\\
B\left[-\frac{1}{\sqrt{h(r)}}\partial_{\psi}I_{\uparrow}+\frac{1}{\sqrt{h(r)}}\frac{\sqrt{3}}{2}
\left(jr^2+2jq-\frac{aq}{2r^2}\right)+\tilde{m}\right]=0
\end{eqnarray}
\begin{eqnarray}\label{Dirac_Goed_4}
\frac{2A}{r\sqrt{\chi(r)}}\left[-\partial_{\theta}I_{\uparrow}-\frac{i}{\sin{\theta}}
\partial_{\varphi}I_{\uparrow}-i\cot{\theta}\partial_{\psi}I_{\uparrow}\right]=0
\end{eqnarray}
The equations (\ref{Dirac_Goed_2}) and (\ref{Dirac_Goed_4}) show
that angular variables can be separated from the radial one. So
the action takes the form (\ref{action_gen_ferm}) again. Then the
pair of equations (\ref{Dirac_Goed_1}) and (\ref{Dirac_Goed_3})
give rise to the following system:
\begin{eqnarray}\label{Dirac_Goed_1_2}
B\left[\frac{2\sqrt{h(r)}}{r\sqrt{V(r)}}E-\frac{\sqrt{V(r)}}{\chi(r)}
R'(r)-\frac{2g(r)}{r\sqrt{h(r)V(r)}}L+\right.\nonumber\\
\left.\frac{2e\sqrt{h(r)}}{r\sqrt{V(r)}}\left(\frac{\sqrt{3}q}{2r^2}-\Phi\right)+
\frac{2eg(r)}{r\sqrt{h(r)V(r)}}\frac{\sqrt{3}}{2}\left(jr^2+2jq-\frac{aq}{2r^2}\right)\right]+\nonumber\\
A\left[\frac{1}{\sqrt{h(r)}}\left(L-\frac{\sqrt{3}}{2}
\left(jr^2+2jq-\frac{aq}{2r^2}\right)\right)+\tilde{m}\right]=0
\end{eqnarray}
\begin{eqnarray}\label{Dirac_Goed_3_2}
A\left[-\frac{2\sqrt{h(r)}}{r\sqrt{V(r)}}E-\frac{\sqrt{V(r)}}{\chi(r)}
R'(r)+\frac{2g(r)}{r\sqrt{h(r)V(r)}}L-\right.\nonumber\\
\left.\frac{2e\sqrt{h(r)}}{r\sqrt{V(r)}}\left(\frac{\sqrt{3}q}{2r^2}-\Phi\right)-
\frac{\sqrt{3}eg(r)}{r\sqrt{h(r)V(r)}}\left(jr^2+2jq-\frac{aq}{2r^2}\right)\right]+\nonumber\\
B\left[-\frac{1}{\sqrt{h(r)}}\left(L-\frac{\sqrt{3}}{2}
\left(jr^2+2jq-\frac{aq}{2r^2}\right)\right)+\tilde{m}\right]=0
\end{eqnarray}
The equations (\ref{Dirac_Goed_2}) and (\ref{Dirac_Goed_4}) are
identical and lead to equation for the angular part of the action.
This equation is the same as for the previous case (\ref{D_eq_2m})
and (\ref{D_eq_4m}):
\begin{equation}
\Theta'(\theta)+\frac{iJ}{\sin{\theta}}-iL\cot{\theta}=0
\end{equation}
So the angular part of the action takes the same form as for the
metric (\ref{Sq_KK_BH}).

In order to get equation for the derivative of the radial part of
the action we make use the same arguments as in previous case. So
we arrive at the equation:
\begin{eqnarray}
R'(r)=\frac{2\chi(r)\sqrt{h(r)}}{rV(r)}\left[\left(E+e\left(\frac{\sqrt{3}q}{2r^2}-\Phi\right)-\frac{g(r)}{h(r)}
\left[L-\frac{e\sqrt{3}}{2}\left(jr^2+2jq-\frac{aq}{2r^2}\right)\right]\right)^2\right.\nonumber\\
\left.+\frac{r^2V(r)}{4h(r)}\left(\tilde{m}^2-\frac{1}{h(r)}\left[L+\frac{\sqrt{3}}{2}\left(jr^2+2jq-\frac{aq}{2r^2}\right)\right]^2
\right)\right]^{1/2}
\end{eqnarray}
Having integrated the latter equation in the vicinity of horizon
point and taking the imaginary part we obtain the relation
(\ref{imag_part_goedel}). So we conclude that temperature for
tunnelling Dirac fermions in case of G\"{o}del universe will be
the same as for scalar particles and takes form
(\ref{temp_Godel_sq}).

\section{Conclusions}
We have investigated tunnelling of scalar particles and Dirac
fermions from the squashed charged rotating black holes in five
dimensional case (\ref{Sq_KK_BH}). The same procedure has been
accomplished for similar type of black hole but in G\"{o}del
universe (\ref{goedel_sq_bh}). To consider tunnelling of scalar
particles we have made use of the Hamilton-Jacobi approach which
is based on the examination of the Hamilton-Jacobi equation
(\ref{HJ_eq_1}). As we noted earlier the Hamilton-Jacobi equation
we used is the quasi-classical limit of the Klein-Gordon equation
(\ref{KG_eq}) that describes scalar particles in quantum
mechanics. To find temperature of a black hole the imaginary part
of the action which satisfies the Hamilton-Jacobi equation should
be found. This fact that the action takes complex values is the
direct consequence of the tunnelling process thorough the horizon.
From the point of view of mathematics complex values for the
action appear due to the integration on the interval which
includes a pole of integrand (the horizon point is the simple pole
for the integrand). The imaginary part of the action allows us to
obtain the Boltzmann factor when we calculate tunnelling
probability (\ref{tunn_prob}). The expressions for temperature of
the black hole that we have obtained here take the same form as it
was obtained by other method
\cite{NakagawaPRD2008,Steltea_PRD_2008}.

To consider tunnelling of Dirac particles the approach proposed by
Kerner and Mann \cite{Kerner_PLB08} has been employed. In case of
fermions we also restrict oneself to the quasi-classical
approximation. For this purpose we have chosen the wave function
of Dirac equation in the form (\ref{wave_func_up}) and supposed
that coefficients $A(t,r,\theta,\phi,\psi)$ and
$B(t,r,\theta,\phi,\psi)$ are constant because we take into
consideration only the terms proportional to $\hbar^{-1}$. From
the written system of equations it follows immediately that
variables can be separated. Then similarly to the case of scalar
particles we have singled out the equation for the derivative of
radial part of the action $I_{\uparrow}$. We note that obtained
expressions  for the radial part of the action in the vicinity of
horizon for Dirac fermions are identical to corresponding
expressions for scalar particles. It should be noticed that
angular part of the action which corresponds to the angle $\th$
can take complex values. But in comparison to the radial part
where imaginary parts for outgoing and ingoing particles take
opposite sign for the angular part $\Theta(\th)$ they take the
same sign and can be cancelled out. So the angular part of the
action does not influence on the determination of temperature. As
a consequence, temperature we have found for tunnelling fermions
is the same as for scalar particles.

\end{document}